\documentclass[12pt]{iopart}
\usepackage{graphicx}
\usepackage{bm}
\usepackage{epsfig}
\usepackage{iopams}
\usepackage{cite}

\begin{document}

\title{The Modulation of de Haas-van Alphen Effect in Graphene by Electric
Field}
\author{Shengli Zhang$^{1,2}$, Ning Ma$^{1}$, Erhu Zhang$^{1 }$ \\
$^{1}$Department of Applied Physics, Xi'an Jiaotong University,\\
Xi'an 710049, China\\
$^{2}$MOE Key Laboratory for Nonequilibrium Synthesis and\\
Modulation of Condensed Matter, Xi'an Jiaotong University,\\
Xi'an 710049, China}
\date{\today }

\begin{abstract}
This paper is to explore the de Haas-van Alphen effect $(dHvA)$
of graphene in the presence of an in-plane uniform electric field. Three
major findings are yielded. First of all, the electric field is found to modulate
the de Haas-van Alphen magnetization and magnetic susceptibility through
the dimensionless parameter $( \beta =\frac{E}{\upsilon _{F}B})$. As
the parameter $\beta $ increases, the values of magnetization
and magnetic susceptibility increase to positive infinity or decrease to negative infinity at
the exotic point $\beta _{c}=1$. Besides, the $dHvA$ oscillation
amplitude rises abruptly to infinity for zero temperature at $\beta _{c}=1$,
but eventually collapses at a finite temperature thereby leading to the
vanishing of de Haas-van Alphen effect. In addition, the magnetic susceptibility depends on the
electric and magnetic fields, suggesting that the graphene should be a
non-linear magnetic medium in the presence of the external field. These
results, which are different from those obtained in the standard
nonrelativistic 2D electron gas, are attributed to its anomalous Landau
level spectrum in graphene.
\end{abstract}

\maketitle

\section{Introduction}

Owing to the progress of experimental methods, graphene (or a
graphite monolayer) is now gaining increasing interests in the field
of physics of electronic systems with reduced dimensionality
\cite{1,2,3,4}. It is promising to be applied in nanoelectronics
because of the exotic chiral features \cite{5,6,7,8,9} in its
electronic structure. In particular, such two-dimensional (2D) or
quasi-two-dimensional systems have led to some of the most startling
discoveries in condensed matter physics in the recent years.
Moreover, these anomalous phenomena are found to be tied to the
remarkable "relativistic-like" spectrum of electrons and holes
in graphene, which makes graphene important and interesting
in physics. One of them that have been experimentally
testified is the abnormality of the 2D quantum Hall effect \cite{10,11}.

Another important physics effect is the de Haas-van Alphen $(dHvA)$
oscillation of the graphene. It has been predicted in Ref.
\cite{12}, which proposes that the magnetization oscillates
periodically in a sawtooth pattern as a function of $1/B$, in
agreement with the old Peierls prediction \cite{13}. A question of
great interest arising here is what will happen if an additional electric field
is applied in graphene. Indeed, the electric and magnetic fields
effects on its magnetization and magnetic susceptibility are of
vital significance to our understanding of the Dirac fermion
behaviors. However, little theoretical or experimental
research has been done on this issue yet.

Motivated by the concerns mentioned above, the present study is to
investigate the 2D $dHvA$ effect of the graphene in
the presence of the electric field. The paper is organized as follows. In section $2$, a brief
introduction is given to the 2D model for graphene. The energy eigenvalues and eigenstates, as well
as the density of states $(DOS)$ and the $dHvA$ oscillation period $
\Delta (1/B)$ are obtained analytically. Section $3$ describes some
details of the magnetization and magnetic susceptibility
study. In the meantime, the analytical expressions are
derived for the magnetization and magnetic susceptibility.
More specifically, the one regarding the condition of $T=0~\mbox{K}$ is reported in
section $3.1$ while the one pertaining to the condition of nonzero temperature in section $
3.2$. The section winds up with a discussion of the modulation of $
dHvA$ in graphene by the electric field. In the last section, the conclusions are presented.

\section{Energy eigenvalues and eigenstates}

In order to investigate the modulation of $dHvA$, we begin
with the study of the energy eigenvalues and eigenstates belonging to
the carriers in graphene. The charge carriers in graphene mimic relativistic
particles with zero rest mass and have an effective `speed of light' $
c^{\ast }=\upsilon _{F}\thickapprox 1.0 \times 10^{6}$
ms$^{-1}$, which is essentially governed by Dirac equation
\cite{10,11}. So we start by considering the Dirac equation for such
a 2D gas of Dirac fermions in crossed electric
$\left[ \vec{E}=\left( -E,0,0\right) ,U =Eex\right] $\ and
magnetic $\left[ \vec{B}=\left( 0,0,B\right) ,\vec{A}=\left(
0,Bx,0\right) \right] $\ fields, where $E$ is the electric field
strength and $B$ the magnetic induction intensity. The
single particle Hamiltonian is then given by
\begin{equation}
\hat{H}=\upsilon _{F}\mathbf{\hat{\alpha}\cdot \Pi }+\mathbf{\hat{I}}eEx
\end{equation}
in which $\hat{\alpha}$\ is the Pauli matrix, $\Pi$ is the canonical
momentum, $\hat{I}$ is the 2 $\times$ 2 unit matrix. Following
Landau and Lifshitz \cite{14} the first-order equation of eigenvalue
problem of $\hat{H}$\ becomes the second-order equation
\begin{equation} \label{2}
\left[ \left( \varepsilon -eEx\right) ^{2}-\left( \upsilon _{F}\vec{p}-e\vec{%
A}\right) ^{2}+e\hbar B\upsilon _{F}^{2}\mathbf{\hat{\alpha} }_{z}+ie\hbar
E\upsilon _{F}\mathbf{\hat{\alpha} }_{x}\right] \Psi =0,
\end{equation}
where $\varepsilon $ is the eigenvalue of $\hat{H}$ and other notations are standard. From Eq. (\ref{2}), we
obtain the energy spectra and eigenfunctions of the problem,
\begin{equation} \label{3}
\varepsilon _{n,k_{y}}=sgn\left( n\right) \sqrt{2\left\vert n\right\vert
\hbar eB}\upsilon _{F}\left( 1-\beta ^{2}\right) ^{3/4}+\hbar \upsilon
_{F}\beta k_{y},
\end{equation}
\begin{equation} \label{4}
\Psi _{n,k_{y}}(x,y)\propto \exp \left( ik_{y}y\right) \exp [-(\beta
/2)\alpha _{y}]\left( _{i^{\left\vert n\right\vert }\phi _{n}\left( \xi
\right) }^{sgn(n)i^{\left\vert n\right\vert -1}\phi _{\left\vert
n\right\vert -1}\left( \xi \right) }\right) ,
\end{equation}
with
\begin{equation}
\xi \equiv \frac{\left( 1-\beta ^{2}\right) ^{1/4}}{l_{c}}\left(
x+l_{c}^{2}k_{y}-sgn(n)\frac{\sqrt{2\left\vert n\right\vert }l_{c}\beta }{%
(1-\beta ^{2})^{1/4}}\right) .
\end{equation}
In Eq. (\ref{3}), $e$ and $\hbar =h/2\pi $ are electron charge and
Planck's constant divided by $2\pi $, respectively. The integer $n$
represents the Landau level index, $k_{y}=2\pi l/L_{y}$ $\left(
l=0,\pm 1,\pm 2,\cdot \cdot \cdot ,\right) $ is the quantum number
corresponding to the translation symmetry along the $y$ axis,
$L_{y}$ stands for the size of the graphene in $y$ direction. The
electric field dependent dimensionless parameter $\beta $ is defined
by $\beta =E/\left( \upsilon _{F}B\right) $ and obeys $\left\vert
\beta\right\vert <1 $, where $\upsilon _{F}$ is the Fermi velocity.
In Eq. (\ref{4}), $\phi _{n}\left( \xi \right) $ are the harmonic
oscillator eigenfunctions.
From Eq. $(5)$ we observe that the centers of the x-dependent orbits are located at
\begin{equation} \label{6}
x_{0}=l_{c}^{2}k_{y}-sgn(n)\frac{\sqrt{2\left\vert n\right\vert }l_{c}\beta
}{(1-\beta ^{2})^{1/4}}
\end{equation}
where $l_{c}=\sqrt{\hbar /eB}$ is the magnetic length.
The eigenvalues of $\hat{H}$ show that the exact energies are given by
the sum of quantized harmonic-oscillator
energies and the potential energy of a charged particle located at
coordinate $x_{0}$ in potential field $U(x)$. They
agree with those in Ref. \cite{15}, the authors of which solved the
problem by transforming the original system into a case with the
null electric field, in terms of a Lorentz boost transformation.

We then count the Landau states $\Psi _{n,k_{y}}$ in the presence of
the potential $U(x)$ following the same argument
employed in the absence of crossed electric field. Since $k_{y}=2\pi
l/L_{y}$, the separation between adjacent allowed $k_{y}$ values is
given by $\delta k_{y}=2\pi /L_{y}$. From Eq. (\ref{6}) we may
relate the possible range $\Delta k_{y}$ of $k_{y}$ to the
physically accessible range $\Delta x_{0}$ of $x_{0}$:
\begin{equation}
\Delta x_{0}=l_{c}^{2}\Delta k_{y}.
\end{equation}
Since $\Delta x_{0}=L_{x}$, in order for the Landau states to be centered
within the strip $0\leq x_{0}\leq L_{x}$\ we must have allowed the range of $
k_{y}$\ values given by $\Delta k_{y}=\Delta x_{0}/l_{c}^{2}.$\ The number
of Landau states $\Psi _{n,k_{y}}\left( x,y\right) $ per unit area for each
quantum number $n$ is:
\begin{equation}
D_{n}=\frac{g_{s}\Delta k_{y}}{L_{x}L_{y}\delta k_{y}}=\frac{2eB}{\hbar \pi }
,
\end{equation}
which is independent of $n$ and the degeneracy yields $g_{s}=4$, accounting
for spin degeneracy and sublattice degeneracy in graphene.

We consider a system of $N$ electrons within an area $S$ moving
in the potential $U(x)$ and the magnetic field $B$. Let the
system remain at $0$ K and accordingly the free energy reduces to the
total energy. The full occupation of Landau levels obeys $
D_{n}S=N/\left( n_{F}+1\right) $ with Fermi quantum number $n_{F}$. The
total energy $E=\sum_{n,k_{y}}\varepsilon _{n,k_{y}}\left( B\right) $ will give a
discontinuous derivative $M=-\partial E/\partial B$ at the field values $
B_{n}$ where $M$ is the magnetization. From $D_{n}S=N/\left( n_{F}+1\right) $
these discontinuities in the magnetization occur at reciprocal fields $
1/B_{n}$, so that the period of magnetization oscillation is given by:
\begin{equation} \label{9}
\Delta \left( \frac{1}{B}\right) =\frac{1}{B_{n+1}}-\frac{1}{B_{n}}=\frac{2e
}{\pi \hbar N_{0}},
\end{equation}
where $B_{n+1}$ and $B_{n}$ are the magnetic induction intensity
corresponding to two neighboring levels, which cross the Fermi level
in succession and $N_{0}=N/S$ is the sheet concentration. Eq. (\ref{9})
means that the discontinuous zero-temperature oscillations are periodic in $1/B$.
It is just compatible with the results regarding the null electric
field obtained by Sharapov et al \cite{12}.

\section{Magnetization and magnetic susceptibility}

\subsection{Zero temperature}

Then we move on to investigate the magnetization of electrons in
graphene in the presence of crossed uniform electric and magnetic
fields at $T=0~\mbox{K}$. For simplicity, we ignore spin-orbit
coupling of electrons in the present
work. The magnetization reads \cite{16,17}
\begin{equation}
M=-\left( \partial E/\partial B\right) _{N}
\end{equation}
where $E$ is the total energy and $N$ denotes the total number of electrons
in graphene. To have units in Tesla, we symbolize $B=\mu_{p0}H$, $\mu_{p0}$ as being
the magnetic permeability of free space, $H$ stands for the magnetic field intensity.

The total energy is given by:
\begin{eqnarray}
E=\sum_{n,k_{y}}\varepsilon _{n,k_{y}} &=& \sum_{n=0}^{\left[ \mu
_{0}^{2}/2e\hbar \upsilon _{F}^{2}B\right] }D_{n}\left( \varepsilon
_{n}-\mu \right) +\mu N_{0} \nonumber \\ &&
 +\sum_{l=-l_{F}}^{l_{y}}\frac{2\pi }{L_{y}}\hbar \upsilon _{F}\beta
l,
 \left( -l_{F}\leq l_{y}\leq
l_{F}\right)
\end{eqnarray}
in which the last term is the additional energy induced by the
electric field
corresponding to the $n_{F}$th level partly occupied by electrons and we refer to it as $E_{add}$. Here $
\varepsilon _{n}$ is defined as $\varepsilon _{n}=\sqrt{2n\hbar eB}\upsilon
_{F}\left( 1-\beta ^{2}\right) ^{3/4}$. The chemical potential\ $\mu =\mu
_{0}\left( 1-\beta ^{2}\right) ^{3/4}$\ is derived analytically where $
\mu _{0}$ refers to the zero-temperature chemical potential (equal
to the Fermi energy) in the absence of electric and magnetic fields
as expressed by $\mu _{0}=\hbar \upsilon _{F}\sqrt{N_{0}\pi }$.
Using the formula for the generalized zeta function \cite{18},
$\zeta \left( z,\upsilon +k\right) =\zeta \left( z,\upsilon \right)
-\sum_{m=0}^{k-1}\left( m+\upsilon \right) ^{-z}$ and $\zeta \left(
-1/2,0\right) \equiv \zeta \left( -1/2\right) \equiv -\left( 1/4\pi
\right) \zeta \left( 3/2\right) $, one can write the first and the
second terms in Eq. $(11) $ as the summation of the regular term,
\begin{equation}
E_{reg}=-\frac{\zeta \left( 3/2\right) \upsilon _{F}}{\pi
^{2}\sqrt{2\hbar }} \left( eB\right) ^{3/2}\left( 1-\beta
^{2}\right) ^{3/4}+\frac{2\mu _{0}^{3} }{3\pi \left( \hbar \upsilon
_{F}\right) ^{2}}\left( 1-\beta ^{2}\right) ^{3/4},
\end{equation}
and the oscillating term
\begin{eqnarray} \label{13}
E_{osc} &=&-\frac{2\sqrt{2\hbar }\upsilon _{F}}{\pi \hbar }\left( eB\right)
^{3/2}\left( 1-\beta ^{2}\right) ^{3/4}\zeta \left( -\frac{1}{2},1+\left[
\frac{\mu _{0}^{2}}{2e\hbar \upsilon _{F}^{2}B}\right] \right)  \nonumber \\
& &-\left( 1-\beta ^{2}\right) ^{3/4}\left[ \frac{2\mu
_{0}^{3}}{3\pi \left( \hbar \upsilon _{F}\right) ^{2}}-\frac{2\mu
_{0}eB}{\pi \hbar }\left(
mod\left[ \frac{\mu _{0}^{2}}{2e\hbar \upsilon _{F}^{2}B}\right] -\frac{1}{2
}\right) \right] .\nonumber \\
\end{eqnarray}
In this expression, $\left[ \mu _{0}^{2}/2e\hbar \upsilon _{F}^{2}B\right] $
stands for the integer part of $\mu _{0}^{2}/2e\hbar \upsilon _{F}^{2}B$ and
the mod $\left[ \mu _{0}^{2}/2e\hbar \upsilon _{F}^{2}B\right] $ is the
fractional part of $\mu _{0}^{2}/2e\hbar \upsilon _{F}^{2}B$.

Making use of $\Gamma $ function $\Gamma \left( n+\alpha \right)
=\int_{0}^{\infty }dss^{n+\alpha -1}e^{-s }$ and Bernoulli
polynomials $B_{n}\left( y\right) $, $\sum_{n=0}^{\infty
}\frac{x^{n}}{n!}B_{n}\left( y\right) =\frac{xe^{xy}}{e^{x}-1},$
$\left( \left\vert x\right\vert <2\pi \right) ,$ we obtain
\begin{eqnarray}
&&\sum_{n=2}^{\infty }\frac{\Gamma \left( n+\alpha \right) B_{n}\left(
y\right) }{n!}x^{n} \nonumber  \\
&=&\int_{0}^{\infty }dss^{\alpha -1}e^{-s}\left[ \frac{sxe^{sxy}}{e^{sx}-1}%
-1-sxB_{1}\left( y\right) \right] ,
\end{eqnarray}
where the explicit expressions for the Bernoulli polynomials $%
B_{0},B_{1},B_{2}$ are%
\begin{equation}
B_{0}\left( y\right) =1, B_{1}\left( y\right) =y-1/2, %
B_{2}\left( y\right) =y^{2}-y-1/6.
\end{equation}
The Bernoulli polynomials depend on $mod\left[ y\right] $ in the
following equations, {\it i.e.} $B_{n}\left( mod\left[ y\right]
\right) .$ For brevity, we write it as $B_{n}\left( y\right) $.

Using the formula \cite{17}
\[
\int_{0}^{\infty }\frac{x^{v-1}e^{-\mu x}dx}{1-e^{-\beta x}}=\frac{1}{\beta
^{v}}\Gamma \left( v\right) \zeta \left( v,\frac{\mu }{\beta }\right) , \left( Re\mu >0,Re v>0\right)
\]
we have
\begin{eqnarray}
E_{osc}=\frac{2\left( eB\upsilon _{F}\right) ^{2}}{\pi ^{3/2}\mu _{0}}\left(
1-\beta ^{2}\right) ^{3/4}\sum_{n=0}^{\infty }\frac{\Gamma \left(
n+1/2\right) B_{n+2}\left( w/2\right) }{\left( n+2\right) !}\left( \frac{%
2e\hbar \upsilon _{F}^{2}B}{\mu _{0}^{2}}\right) ^{n} \nonumber \\
\end{eqnarray}
in which $w=\mu _{0}^{2}/\left( e\hbar \upsilon _{F}^{2}B\right) $.
Regarding small fields, $eB\hbar \upsilon _{F}^{2}\ll \mu _{0}^{2}$, we can
apply the following asymptotic expansions for $J_{1}=\int_{0}^{\infty
}dte^{-tp}/\left[ \sqrt{\pi t}\left( t^{2}+1\right) \right] $:%
\begin{equation}
J_{1}\left( p\right) =\frac{1}{\sqrt{\pi }}\sum_{n=0}^{\infty }\frac{\left(
-1\right) ^{n}\Gamma \left( n+1/2\right) }{p^{n+1/2}}
\end{equation}
and the Bernoulli polynomials $B_{n}$ periodically continue beyond the
interval [0,1]:
\begin{eqnarray}
B_{n} &=&-\frac{2n!}{\left( 2\pi \right) ^{n}}\sum_{k=1}^{\infty }\frac{1}{%
k^{n}}\cos \left( 2\pi k x-\frac{n\pi }{2}\right) ,  \\
n &>&1,0\leq x\leq 1; n=1,0<x<1.  \nonumber
\end{eqnarray}
It is easy to get the following form:
\begin{equation} \label{19}
E_{osc}=\frac{\left( eB\right) ^{3/2}\upsilon _{F}}{\sqrt{\hbar }\pi }\left(
1-\beta ^{2}\right) ^{3/4}\sum_{k=1}^{\infty }\frac{1}{\left( \pi k\right)
^{3/2}}J_{1}\left( \pi kw\right) \cos \left( \pi kw\right) .
\end{equation}
For $\sqrt{eB\hbar \upsilon _{F}^{2}}\ll \mu _{0}$, keeping the leading term
in asymptotic expansions for $J_{1}\left( p\right) $, we finally obtain from
Eq. (\ref{19})
\begin{equation}
E_{osc}\cong \frac{\left( eB\upsilon _{F}\right) ^{2}}{\pi \mu _{0}}\left(
1-\beta ^{2}\right) ^{3/4}\sum_{k=1}^{\infty }\frac{\cos \left( \pi
kw\right) }{\left( \pi k\right) ^{2}}.
\end{equation}
Hence the total energy $E$ can be expressed as a sum of regular, oscillating
and the additional energy terms,
\begin{equation}
E=E_{reg}+E_{osc}+E_{add}.
\end{equation}

According to the results reported above, we get the corresponding de
Haas-van Alphen magnetization,
\begin{equation} \label{22}
M_{reg}=\frac{3e\sqrt{eB\hbar }\varsigma \left( 3/2\right) \upsilon _{F}}{2%
\sqrt{2}\pi ^{2}\hbar \left( 1-\beta ^{2}\right) ^{1/4}}-\frac{\mu
_{0}^{3}\beta ^{2}}{\pi \left( \hbar \upsilon _{F}\right) ^{2}B\left(
1-\beta ^{2}\right) ^{1/4}},
\end{equation}
\begin{equation}
M_{add}=\frac{\pi \hbar \upsilon _{F}\beta }{BL_{y}}A_{M_{0}},
\end{equation}
where
\begin{equation} \label{24}
A_{M_{0}}=\left[ \frac{\mu _{0}^{4}}{\pi ^{2}\left( \hbar \upsilon _{F}\right)
^{4}}-\left[ \frac{\mu _{0}^{2}}{2e\hbar \upsilon _{F}^{2}B}\right] \left(
\left[ \frac{\mu _{0}^{2}}{2e\hbar \upsilon _{F}^{2}B}\right] +1\right)
\left( \frac{2eB}{\hbar \pi }\right) ^{2}\right].
\end{equation}
This expression involves a dependence on the integer part $\left[ \mu
_{0}^{2}/2e\hbar \upsilon _{F}^{2}B\right] $ and
\begin{eqnarray} \label{25}
M_{osc}=A_{M1}\sum_{k=1}^{\infty }\frac{(-1)^{k}}{\pi k}\sin \left( \frac{
\pi k\mu _{0}^{2}}{e\hbar \upsilon _{F}^{2}B}\right)
+A_{M2}\sum_{k=1}^{\infty }\frac{(-1)^{k}}{\left( \pi k\right) ^{2}}\cos
\left( \frac{\pi k\mu _{0}^{2}}{e\hbar \upsilon _{F}^{2}B}\right) , \nonumber \\
\end{eqnarray}
where $A_{M1}=-\frac{e\mu _{0}\left( 1-\beta ^{2}\right) ^{3/4}}{\pi \hbar }$
and $A_{M2}=-\frac{\left( e\upsilon _{F}\right) ^{2}B\left( 4-\beta
^{2}\right) }{2\pi \mu _{0}\left( 1-\beta ^{2}\right) ^{1/4}}$.

Figure 1 shows that the $dHvA$\ oscillation on the magnetization $M$
modulated by electric field in graphene. As shown in Fig. 1(a), the magnetization $M$
oscillates periodically in $1/B$ with the period of $\Delta \left( 1/B\right) =2e\hbar \upsilon _{F}^{2}/\mu
_{0}^{2}$. Three oscillation curves correspond to the applied electric
field strength $E_{1}= 5$ V/m, $E_{2}= 10$ V/m and $E_{3}= 15$ V/m,
respectively. From these curves, one can see that the oscillation
amplitude $(OA)$ of magnetization is proportional to the electric field $
E$. Also, the $OA$ ascends significantly with increasing $1/B$, but
remains unchanged for the null electric field in Ref. \cite{12}.
Thus, it is demonstrated the electric field effect on the $OA$ of
magnetization. However unexpected it may be, according to Eq. (\ref{25})
the $OA$ increases abruptly
to infinity at $\beta _{c}= 1$ thereby leading to the vanishing of $
dHvA$\ oscillation. It is interesting that the abnormal phenomena will die out if
the electric filed vanishes. Thus, it can be also attributed to the electric
field effect. We hope that the new findings will be verified when magnetization
experiments under in-plane electric field are carried out in graphene. The predicted
effect will hopefully also help interpretation of magnetization in experiment.

Meanwhile, we find that these peculiar features are
absent in standard quantum 2D electron gas systems \cite{19,20,21}.
Hence, the possible reason for the effect might be that it is
determined by the "relativistic" character of carriers in graphene,
unlike the usual sample, which can be traced to the exotic structure
in graphene.

\begin{figure}
  \begin{center}
    \includegraphics[width=14cm]{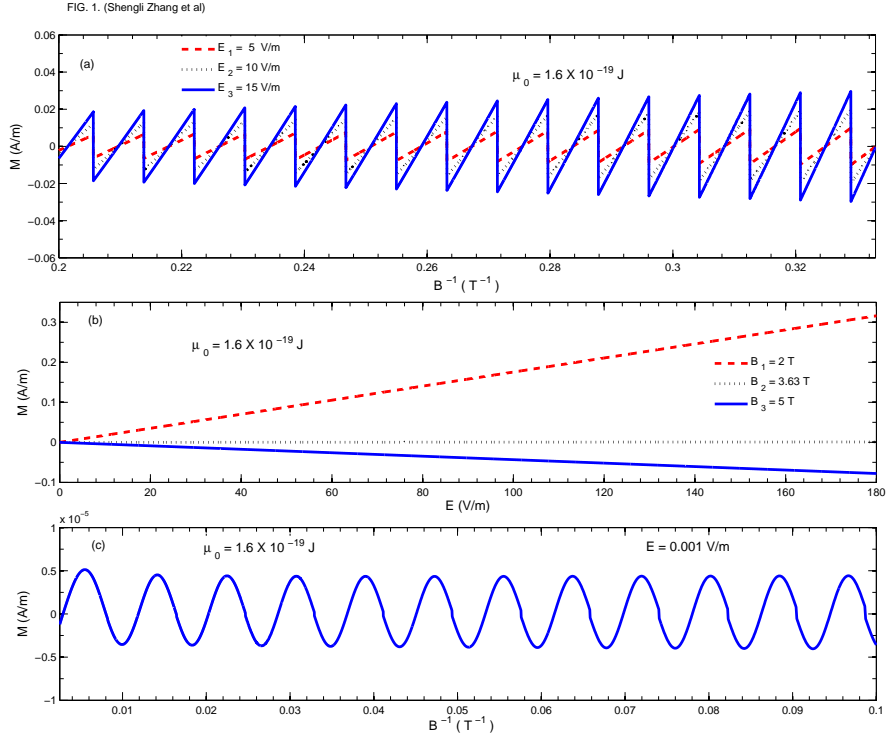}\\
  \caption{$\left( a\right) $ Magnetization $M$ is plotted as a function of
reciprocal magnetic field $1/B$ for a given chemical potential $\mu
_{0}=1.6\times 10^{-19}~\mbox{J}$ and $T=0~\mbox{K}$. Three oscillation curves correspond to
$E_{1}=5~\mbox{V/m}$, $E_{2}=10~\mbox{V/m}$ and $E_{3}=15~\mbox{V/m}$, respectively. $\left( b\right) $
$M$ plotted vs electric field $E$ for $\mu_{0}=1.6\times 10^{-19}~\mbox{J}$ and $T=0~\mbox{K}$. Three curves correspond
to $B_{1}=2~\mbox{T}$, $B_{2}=3.63~\mbox{T}$ and $B_{3}=5~\mbox{T}$, respectively. $\left( c\right) $ The magnetization, $M$,
as a function of $1/B$ at $E=0.001~\mbox{V/m}$, $\mu
_{0}=1.6\times 10^{-19}~\mbox{J}$ and $T=0~\mbox{K}$.}\label{1}
  \end{center}

\end{figure}

Figure 1(b) demonstrates that the magnetization $M$ is a function
of the electric field $E$. Three curves correspond to $
B_{1}=2~\mbox{T}$, $B_{2}=3.63~\mbox{T}$ and $B_{3}=5~\mbox{T}$, respectively. They exhibit the
magnetization varies approximately linearly with increasing $E$ within the given values of parameters.
For $B_{1}=2~\mbox{T}$, the magnetization increases with increasing E. But for $B_{3}=5~\mbox{T}$, as $E$ increases
the magnetization decreases. Especially, there is a special behavior of the magnetization at some magnetic fields ({\it e.g.} $
B_{2}=3.63~\mbox{T}$). In this case, the magnetization satisfies $M\simeq 0$, accounting for the
disappearance of magnetization in graphene.
 Note that, the magnetization become infinite when the variation
of $E$ obeys $\beta _{c}=1$ according to Eqs. (\ref{22}) and (\ref{25}).
For example, the dashed line $(B_{1}=2~\mbox{T})$
rises abruptly to infinity at $E=2\times 10^{6}~\mbox{V/m}$, strikingly
different from the non-relativistic results.

In Fig. 1(c), we present the $dHvA$ effect for a wide range of
magnetic fields starting from 10 to 400 T, in order to examine what
will happen if the magnetic field tends to be infinite. As a result, we
observe that on this scale, the electric field effect could be
negligible corresponding to the case $\beta\rightarrow 0$. In
other words, for the case of $\beta\rightarrow 0$, the magnetization
$M$ reduces to the result for the null electric field.

The same discussion above fits for the magnetic susceptibility $\chi $.
We can obtain the expression of the $dHvA$ magnetic susceptibility in terms
of $\chi =\partial M/\partial H$,
\begin{equation} \label{26}
\chi _{reg}=\frac{3e\mu_{p0}\sqrt{e\hbar }\varsigma \left( 3/2\right) \upsilon _{F}}{%
4\sqrt{2B}\pi ^{2}\hbar \left( 1-\beta ^{2}\right) ^{5/4}}\left( 1-2\beta
^{2}\right) +\frac{\mu _{0}^{3}\beta ^{2}\mu_{p0}\left( 6-5\beta ^{2}\right) }{2\pi
\left( \hbar \upsilon _{F}B\right) ^{2}\left( 1-\beta ^{2}\right) ^{5/4}},
\end{equation}
\begin{equation}
\chi _{add}=-\frac{2\mu_{p0}\beta \mu _{0}^{4}}{L_{y}B^{2}\left( \hbar \upsilon
_{F}\right) ^{3}},
\end{equation}
and
 \begin{eqnarray} \label{28}
\chi _{osc}&=&\left( A_{\chi 1}+A_{\chi 2}\right) \sum_{k=1}^{\infty
}(-1)^{k}\cos \left( \frac{\pi k\mu _{0}^{2}}{e\hbar \upsilon _{F}^{2}B}%
\right)  \nonumber \\ &&
 +(A_{\chi 3}+A_{\chi 4})\sum_{k=1}^{\infty }\frac{(-1)^{k}}{\pi k}%
\sin \left[ \frac{\pi k\mu _{0}^{2}}{e\hbar \upsilon _{F}^{2}B}\right] ,
\end{eqnarray}
where
\begin{eqnarray}
A_{\chi 1} &=&\frac{\mu_{p0}\mu _{0}^{3}\left( 1-\beta ^{2}\right) ^{3/4}}{\pi
\left( \hbar \upsilon _{F}B\right) ^{2}}, A_{\chi 2}=-\frac{\left(
e\upsilon _{F}\right) ^{2}\mu_{p0}\left( 8-10\beta ^{2}-\beta ^{4}\right) }{4\pi \mu
_{0}\left( \pi k\right) ^{2}\left( 1-\beta ^{2}\right) ^{5/4}}, \nonumber  \\
A_{\chi 3} &=&-\frac{\mu_{p0}e\mu _{0}\left( 4-\beta ^{2}\right)
}{2\pi \hbar B\left( 1-\beta ^{2}\right) ^{1/4}}, A_{\chi
4}=-\frac{3\mu_{p0}e\mu _{0}\beta ^{2}}{2\pi \hbar B\left( 1-\beta
^{2}\right) ^{1/4}}.
\end{eqnarray}
We can see that the magnetic susceptibility
is related to both the electric and magnetic fields. It follows that, unlike
the usual samples, graphene may be a non-linear magnetic medium.

\begin{figure}
  \begin{center}
  \includegraphics[width=14cm]{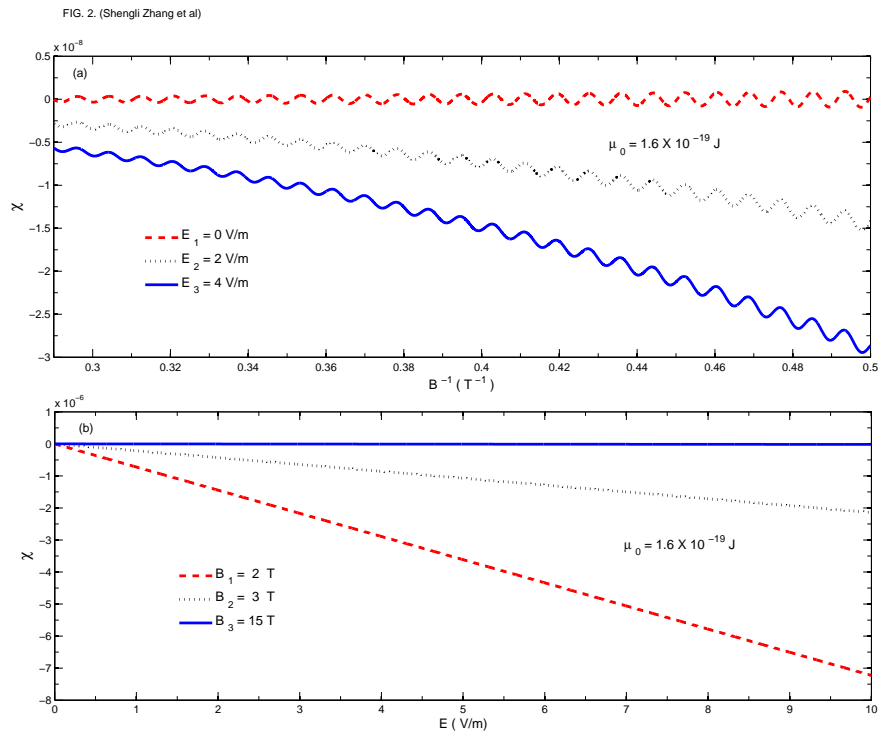}\\
  \caption{$\left( a\right) $ Magnetic susceptibility $\chi $ depends on reciprocal
field $1/B$ for a given chemical potential $\mu_{0}=1.6\times 10^{-19}~\mbox{J}$ and $T=0~\mbox{K}$.
Three oscillation curves of magnetic susceptibility correspond to $
E_{1}=0~\mbox{V/m}$, $E_{2}=2~\mbox{V/m}$ and $E_{3}=4~\mbox{V/m}$, respectively. $\left( b\right) $
$\chi $ as a function of electric field $E$ for $\mu _{0}=1.6\times 10^{-19}~\mbox{J}$ and $T=0~\mbox{K}$. Three curves
correspond to $B_{1}=$ 2 T, $B_{2}=$ 3 T and $B_{3}=$ 15 T, respectively.
}\label{2}
  \end{center}

\end{figure}

Figure 2(a) shows that the magnetic susceptibility $\chi $ oscillates periodically
as a function of $1/B$ and the period follows Eq. (\ref{9}). Three
oscillation curves correspond to $E_{1}=0~\mbox{V/m}$, $
E_{2}=2~\mbox{V/m}$ and $E_{3}=4~\mbox{V/m}$, respectively. For the
case $(E_{1})$, the reader may observe that the magnetic
susceptibility $\chi $ swings between negative and positive values,
thus the curve shows a totally orbital diamagnetic to paramagnetic
transition. As for the
finite electric fields, one can see that the magnetic susceptibility
decreases in company with the periodic oscillation while $1/B$
rising and the $OA$ augments as $1/B$ increases. Furthermore,
from Eq. (\ref{28}) it can be seen that the $OA$ increases to infinity at $
\beta _{c}=1$, leading to the vanishing of $dHvA$ effect on
magnetic susceptibility. In general, the magnetic susceptibility $\chi $ is a constant in
usual electron gas, but in graphene it exhibits the dependence on the
external field.

Figure 2(b) depicts the magnetic susceptibility $\chi $ with respect
to the electric field E. Three curves correspond to $B_{1}=2~\mbox{T}$, $B_{2}=3~\mbox{T}$ and
$B_{3}=15~\mbox{T}$, respectively. As is shown by the three
curves, the magnetic susceptibility varies approximately linearly
with increasing $E$. For $B= 2$ T, or 3 T, the magnetic
susceptibility decreases as $E$ increases and yields $\chi < 0$,
indicating the existence of Landau diamagnetism in graphene, the
origin of which can be traced to the quantized Landau level. In the
case of $B=15~\mbox{T}$ or larger, the magnetic susceptibility $\chi\simeq 0$ despite the increase of $E$,
suggesting the disappearance of the diamagnetism in graphene. That
is, there is no increase in the magnetic susceptibility with
increasing $E$. Moreover, the dashed line
$(B_{1}=2~\mbox{T}$) decreases to negative infinity at the exotic point
$E=2\times 10^{6}$ V/m ({\it i.e.} $\beta _{c}=1$) as
illustrated by Eqs. (\ref{26}) and (\ref{28}).

\subsection{Finite temperature}

We now consider the temperature effect on the oscillations of
magnetization and magnetic susceptibility. As documented in \cite{12},
the thermodynamic potential of electrons in graphene, can be
expressed as
\begin{equation} \label{30}
\Omega \left( T,\mu \right) =\int_{-\infty }^{\infty }d\omega P_{T}\left(
\omega -\mu \right) E\left( \omega \right) ,
\end{equation}
with energy variable $\omega $ and $\mu $. And $P_{T}\left( z\right) $ is the
distribution function as
\begin{equation}
P_{T}\left( z\right) =-\frac{\partial n_{F}\left( z\right) }{\partial z}=
\frac{1}{4k_{B}T\cosh ^{2}\frac{z}{2k_{B}T}}.
\end{equation}

Using Eq. $\left( 30\right) $, the thermodynamic potential $
\Omega $ can be divided as follows:
\begin{equation}
\Omega \left( T,\mu \right) =\Omega _{reg}+\Omega _{add}+\Omega _{osc}.
\end{equation}
At the low temperatures, neglecting $0(k_{B}T),$ we can obtain
\begin{equation}
\Omega _{reg}=-\frac{\zeta \left( 3/2\right) \upsilon _{F}}{\pi ^{2}\sqrt{%
2\hbar }}\left( eB\right) ^{3/2}\left( 1-\beta ^{2}\right) ^{3/4}+\frac{2\mu
_{0}^{3}}{3\pi \left( \hbar \upsilon _{F}\right) ^{2}}\left( 1-\beta
^{2}\right) ^{3/4},
\end{equation}
\begin{equation}
\Omega _{add}=\frac{\pi }{L_{y}}\hbar \upsilon _{F}\beta A_{\Omega
_{T}}R_{T}\left( k,\mu \right) ,
\end{equation}
where
\begin{eqnarray} \label{35}
A_{\Omega _{T}} &=&\frac{\mu _{0}^{4}}{\pi ^{2}\left( \hbar \upsilon
_{F}\right) ^{4}}-\frac{2eB\mu _{0}^{2}}{\hbar \left( \pi \hbar \upsilon
_{F}\right) ^{2}}(2\left[ \frac{\mu _{0}^{2}}{2e\hbar \upsilon _{F}^{2}B}
\right] +1)+\left[ \frac{\mu _{0}^{2}}{2e\hbar \upsilon _{F}^{2}B}\right] \nonumber \\
&&\times \left( \left[ \frac{\mu _{0}^{2}}{2e\hbar \upsilon _{F}^{2}B}\right]
+1\right) \left( \frac{2eB}{\hbar \pi }\right) ^{2}.
\end{eqnarray}
This expression involves a dependence on the integer part $[\mu
_{0}^{2}/2e\hbar \upsilon _{F}^{2}B]$. And here we introduced the
temperature reduction factor
\begin{equation} \label{36}
R_{T}\left( k,\mu \right) =\frac{2\pi ^{2}k\mu _{0}k_{B}T/\left( e\hbar
\upsilon _{F}^{2}B^{\ast }\right) }{\sinh \frac{2\pi ^{2}k\mu _{0}k_{B}T}{
e\hbar \upsilon _{F}^{2}B^{\ast }}},
\end{equation}
in which $B^{\ast }=B\left( 1-\beta
^{2}\right) ^{3/4}$. Eq. (\ref{36}) means that the temperature reduction factor $
R_{T}$ depends not only on the temperature, but also on the electric and
magnetic fields. We return now to the oscillating part of thermodynamic
potential. Substituting Eq. (\ref{19}) into Eq. (\ref{30}), it is convenient to
get the expression of $\Omega _{osc}$ as
\begin{eqnarray}
\Omega _{osc}\left( T,\mu \right) &=&\frac{\left( eB\right) ^{3/2}\upsilon
_{F}}{\sqrt{\hbar }\pi ^{3/2}}\left( 1-\beta ^{2}\right)
^{3/4}\sum_{k=1}^{\infty }\frac{1}{\left( \pi k\right) ^{3/2}} \nonumber \\
&&\times Im [ e^{-i\pi /4}\int_{0}^{\infty }\frac{dt
e^{-i\left( \pi kw\right) t}}{\sqrt{t}\left( t+1\right) }\int_{-\infty
}^{\infty }\frac{d\epsilon }{4k_{B}T\cosh ^{2}\frac{\left( \epsilon -\mu
\right) }{2k_{B}T}} \nonumber \\
&& \exp \left( \frac{-i\pi k\epsilon ^{2}\left( t+1\right) }{
e\hbar \upsilon _{F}^{2}B\left( 1-\beta ^{2}\right) ^{3/2}}\right) ] ,
\end{eqnarray}
in which we used that the function $J_{1}(p,r)$ and $J_{2}(p,r)$ which can be
represented as -Im and Re parts of the same function
\begin{eqnarray}
\sqrt{\pi }J_{1}\left( p,r\right) &=&-Im \int_{0}^{\infty }\frac{dt
 e^{-pt-r/t}}{\sqrt{t}\left( t+i\right) }, \nonumber \\
\sqrt{\pi }J_{2}\left( p,r\right) &=&-Re\int_{0}^{\infty }\frac{dt
 e^{-pt-r/t}}{\sqrt{t}\left( t+i\right) },
\end{eqnarray}
and rotated the integration contour to the imaginary axis.
Finally we get
\begin{equation} \label{39}
\Omega _{osc}\left( T,\mu \right) =\frac{\left( eB\upsilon _{F}\right) ^{2}}{
\pi \mu _{0}}\left( 1-\beta ^{2}\right) ^{3/4}\sum_{k=1}^{\infty }\frac{\cos
\left( \pi kw\right) }{\left( \pi k\right) ^{2}}R_{T}\left( k,\mu \right) ,
\end{equation}
with $w=\mu _{0}^{2}/\left( e\hbar \upsilon _{F}^{2}B\right) $. Clearly,
since $R_{T}\left( k,\mu \right) \rightarrow 1$ for $T\rightarrow 0,$ Eq. (\ref{39})
reduces to the oscillating energy for zero temperature.
When the temperature $T\neq 0,$ the magnetization can be obtained as follows:
\begin{equation} \label{40}
M_{reg}^{T}=\frac{3e\sqrt{eB\hbar }\zeta \left( 3/2\right) \upsilon _{F}}{2
\sqrt{2}\pi ^{2}\hbar \left( 1-\beta ^{2}\right) ^{1/4}}-\frac{\mu
_{0}^{3}\beta ^{2}}{\pi \left( \hbar \upsilon _{F}\right) ^{2}B\left(
1-\beta ^{2}\right) ^{1/4}},
\end{equation}
\begin{equation} \label{41}
M_{add}^{T}=\frac{\pi \hbar \upsilon _{F}\beta R_{T}}{BL_{y}}A_{M_{0}}-\frac{
\pi \hbar \upsilon _{F}\beta \left( 2+\beta ^{2}\right) }{2BL_{y}\left(
1-\beta ^{2}\right) }A_{\Omega _{T}}\left( R_{T}^{\ast }-R_{T}\right) ,
\end{equation}
and
\begin{eqnarray} \label{42}
M_{osc}^{T} &=&-\frac{e\mu _{0}}{\pi \hbar }\left( 1-\beta ^{2}\right)
^{3/4}\sum_{k=1}^{\infty }\frac{\sin \left( \pi kw\right) }{\pi k}R_{T}-
\frac{\left( e\upsilon _{F}\right) ^{2}B}{2\pi \mu _{0}\left( 1-\beta
^{2}\right) ^{1/4}} \nonumber \\
&&\times \lbrack \left( 4-\beta ^{2}\right) R_{T}+\left( 2+\beta ^{2}\right)
\left( R_{T}^{\ast }-R_{T}\right) ]\sum_{k=1}^{\infty }\frac{\cos \left( \pi
kw\right) }{\left( \pi k\right) ^{2}},
\end{eqnarray}
where $R_{T}^{\ast }\left( k,\mu \right) =R_{T}^{2}\left( k,\mu \right)
\cosh \left( 2\pi ^{2}k\mu _{0}k_{B}T/e\hbar \upsilon _{F}^{2}B^{\ast
}\right) $. Apparently, since $R_{T}\left( k,\mu \right) \rightarrow 1$ and $
R_{T}^{\ast }\left( k,\mu \right) \rightarrow 1$ for $T\rightarrow 0,$ Eq. (\ref{42})
reduces to the magnetization in Eq. (\ref{25}) for zero temperature.

\begin{figure}
  \begin{center}
  \includegraphics[width=14cm]{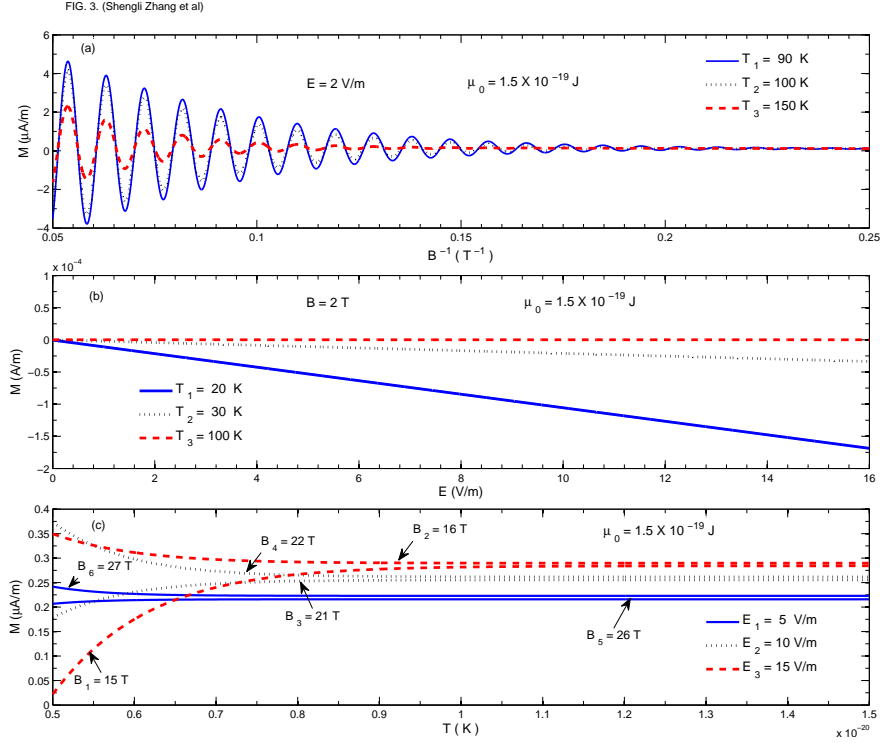}\\
  \caption{$\left( a\right) $\ Magnetization $M$ is plotted as a function of
reciprocal magnetic field $1/B$ for a given chemical potential $\mu
_{0}=1.5\times 10^{-19}~\mbox{J}$ and $E=2~\mbox{V/m}$. Three curves correspond to
temperature $T_{1}=90~\mbox{K}$, $T_{2}=100~\mbox{K}$ and $T_{3}=150~\mbox{K}$, respectively. $\left( b\right) $
\ $M$\ plotted to electric field $E$ with $B=2$ T and $\mu _{0}=1.5\times 10^{-19}~\mbox{J}$. Three curves correspond
to $T_{1}=20~\mbox{K}$, $T_{2}=30~\mbox{K}$ and $T_{3}=100~\mbox{K}$, respectively. $\left( c\right) $ $M$ is
plotted as a function of temperature $T$ for $\mu _{0}=1.5\times 10^{-19}~\mbox{J}$.

}\label{3}
  \end{center}

\end{figure}

As is illustrated in Figure 3(a), finite $T$ causes a reduction of the
magnetization amplitude as opposed to the case of $T=0~\mbox{K}$. Meanwhile, it
shows the magnetization $M$ versus reciprocal field $1/B$ for three different
temperatures with a given $E$. It can be seen that as the
value of $1/B$ increases, the $OA$ of magnetization decreases and eventually
collapses when $\beta\rightarrow 1$. Nevertheless, regarding the zero
temperature, there is no such a collapse for the $OA$ of magnetization as
shown in Fig. $1(a)$. Accordingly, we attribute them to the finite temperature
effect.

Figure 3(b) depicts the magnetization $M$ as a function of $E$
for three different temperatures with a given $B$. Three curves
correspond to $T_{1}=20~\mbox{K}$, $T_{2}=30~\mbox{K}$ and $
T_{3}=100~\mbox{K}$, respectively. It has been shown that the magnetization
decreases approximately linearly with increasing $E$ except in the case
of $T_{3}=100~\mbox{K}$. Finally, the magnetization decreases to the negative
infinity at $\beta _{c}=1$ following Eqs. (\ref{40}), (\ref{41}) and (\ref{42}). At the
temperature $T_{3}$ or even a higher one, the magnetization $M\simeq 0$
rather than decrease with increasing $E$. Fig. $3(c)$ gives the
magnetization $M$ as function of the temperature $T$. Increasing $T$, the
magnetization $M$ increases for $B_{1}$, $B_{3}$, $B_{5}$ and decreases for $
B_{2}$, $B_{4}$, $B_{6}$, but they all end up approaching different
constants.

We get the $dHvA$\ magnetic susceptibility from the magnetization:
\begin{equation} \label{43}
\chi _{reg}^{T}=\frac{3e\mu_{p0}\sqrt{e\hbar }\varsigma \left( 3/2\right) \upsilon
_{F}}{4\sqrt{2B}\pi ^{2}\hbar \left( 1-\beta ^{2}\right) ^{5/4}}\left(
1-2\beta ^{2}\right) +\frac{\mu _{0}^{3}\beta ^{2}\mu_{p0}\left( 6-5\beta
^{2}\right) }{2\pi \left( \hbar \upsilon _{F}B\right) ^{2}\left( 1-\beta
^{2}\right) ^{5/4}},
\end{equation}
\begin{eqnarray} \label{44}
\chi _{add}^{T}&=&-\frac{2\mu_{p0}\beta \mu
_{0}^{4}R_{T}}{L_{y}B^{2}\left( \hbar \upsilon _{F}\right)
^{3}}+\frac{\pi \hbar \upsilon _{F}\mu_{p0}\beta \left( 2+\beta
^{2}\right) }{B^{2}L_{y}\left( 1-\beta ^{2}\right) }A_{M_{0}}\left(
R_{T}^{\ast }-R_{T}\right) \nonumber \\ && -A_{\chi 0_{T}}A_{\Omega
_{T}}\frac{\pi \hbar \upsilon _{F}\mu_{p0}\beta }{L_{y}},
\end{eqnarray}
and
\begin{equation} \label{45}
\chi _{osc}^{T}=A_{\chi 1_{T}}\sum_{k=1}^{\infty }\cos \left( \pi kw\right)
+A_{\chi 2_{T}}\sum_{k=1}^{\infty }\sin \left( \pi kw\right),
\end{equation}
in Eq. (\ref{44}), $A_{M_{0}}$ and $A_{\Omega _{T}}$ are defined
by Eqs. (\ref{24}) and (\ref{35}), respectively.
We also define,
\begin{eqnarray}
A_{\chi 0_{T}} &=&\frac{\left( 2\beta ^{4}-10\beta ^{2}-4\right) }{\left[
2B\left( 1-\beta ^{2}\right) \right] ^{2}}[R_{T}^{\ast }-R_{T}]+\left( \frac{
2+\beta ^{2}}{2B\left( 1-\beta ^{2}\right) }\right) ^{2}[2R_{T} \nonumber \\
&&\times (R_{T}^{\ast }-R_{T})\cosh \frac{2\pi ^{2}kT\mu
_{0}}{e\hbar
\upsilon _{F}^{2}B^{\ast }}-R_{T}^{3}\sinh ^{2}\frac{2\pi ^{2}kT\mu _{0}}{
e\hbar \upsilon _{F}^{2}B^{\ast }}-\left( R_{T}^{\ast }-R_{T}\right)
], \nonumber \\
\end{eqnarray}
\begin{eqnarray}
A_{\chi 1_{T}} &=&\frac{\mu _{0}^{3}\mu_{p0}\left( 1-\beta ^{2}\right) ^{3/4}}{\pi
\left( \hbar \upsilon _{F}B\right) ^{2}}R_{T}-\frac{\left( e\upsilon
_{F}\right) ^{2}\mu_{p0}\left( 4-\beta ^{2}\right) \left( 2+\beta ^{2}\right) }{2\pi
\mu _{0}\left( \pi k\right) ^{2}\left( 1-\beta ^{2}\right) ^{5/4}}\left(
R_{T}^{\ast }-R_{T}\right) \nonumber \\
&&-R_{T}\frac{\left( e\upsilon _{F}\right) ^{2}\mu_{p0}(8-10\beta ^{2}-\beta ^{4})}{
4\pi \mu _{0}\left( \pi k\right) ^{2}\left( 1-\beta ^{2}\right) ^{5/4}}-
\frac{A_{\chi 0_{T}}\mu_{p0}\left( e\upsilon _{F}B\right) ^{2}}{\pi
\mu _{0}\left( \pi k\right) ^{2}}\left( 1-\beta ^{2}\right) ^{3/4},
\nonumber \\
\end{eqnarray}
\begin{equation}
A_{\chi 2_{T}}=-\frac{e\mu _{0}\mu_{p0}\left( 2+\beta ^{2}\right) }{\pi ^{2}\hbar
kB\left( 1-\beta ^{2}\right) ^{1/4}}R_{T}^{\ast }.
\end{equation}

\begin{figure}
  \begin{center}
  \includegraphics[width=14cm]{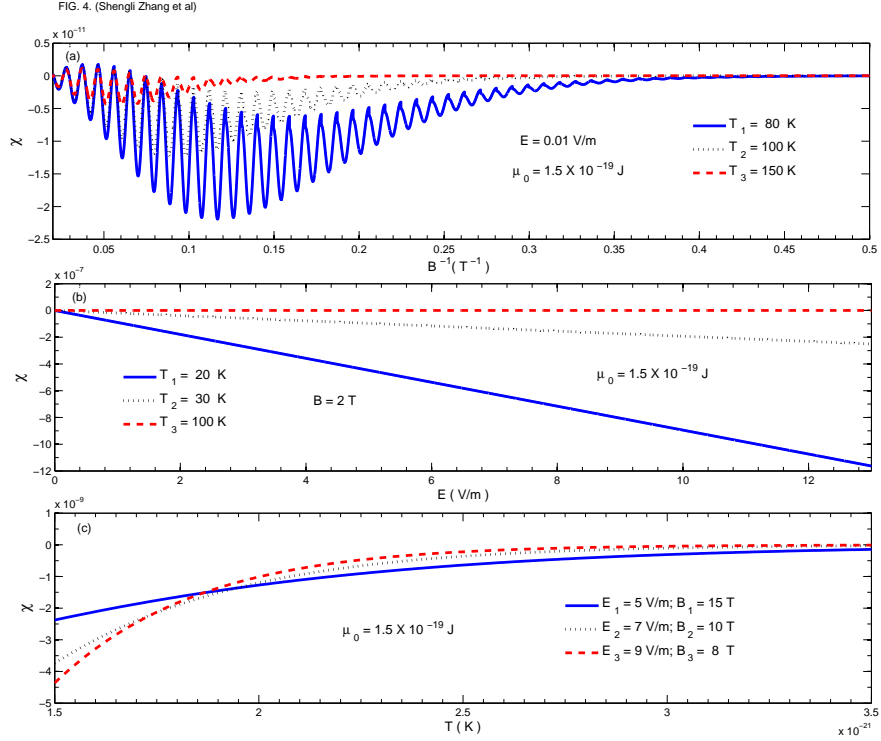}\\
  \caption{$\left( a\right) $ Magnetic susceptibility $\chi $ to reciprocal field $1/B$
for a given chemical potential $\mu _{0}=1.5\times 10^{-19}~\mbox{J}$ and $
E=0.01~\mbox{V/m}$. Three oscillation curves of magnetic susceptibility correspond to $
T_{1}=80~\mbox{K}$, $T_{2}=100~\mbox{K}$ and $T_{3}=150~\mbox{K}$, respectively. $\left( b\right) $ $\chi $
as a function of electric field $E$ with $B=$ 2 T and $\mu _{0}=1.5\times 10^{-19}~\mbox{J}$. Three oscillation curves of
magnetic susceptibility correspond to $T_{1}=20~\mbox{K}$, $T_{2}=30~\mbox{K}$ and $T_{3}=100~\mbox{K}$,
respectively. $\left( c\right) $ $\chi $ as a function of temperature $T$
for $\mu _{0}=1.5\times 10^{-19}~\mbox{J}$.

}\label{4}
  \end{center}

\end{figure}

Figure \ref{4}(a) shows the magnetic susceptibility $\chi $ oscillates periodically as a
function of $1/B$, for three characteristic temperatures, namely, 80, 100, and 150 K. Similar
to the magnetization as shown in Fig. $3(a)$, it also exhibits the dependence on temperature. It can be seen clearly that
finite $T$ causes a reduction of the oscillation amplitude and as the
value of $1/B$ increases, the $OA$ of magnetic susceptibility finally decays to zero.

Figure \ref{4}(b) shows the magnetic susceptibility $\chi $ plotted as a function of $E$
with a given $B$. Three curves correspond to $
T_{1}=20~\mbox{K}$, $T_{2}=30~\mbox{K}$ and $T_{3}=100~\mbox{K}$, respectively. For a finite
temperature such as $T_{1}$ or $T_{2}$, the magnetic susceptibility decreases
approximately linearly with increasing $E$ and exhibits the Landau
diamagnetism in graphene supported by $\chi<0$. In contrast, at the
temperature $T_{3}$, the magnetic susceptibility $\chi\simeq 0$, standing for the
disappearance of the diamagnetism in graphene. In addition, there is no
increase in the magnetic susceptibility with increasing $E$. It warrants great
attention that from Eqs. (\ref{43}), (\ref{44}) and (\ref{45}), it could be inferred
that the magnetic susceptibility eventually decreases to the negative infinity at $
\beta _{c}=$ 1. Fig. $4(c)$ directly gives the magnetic susceptibility $\chi $ with
respect to the temperature $T$. More specifically, it shows that the
magnetic susceptibility $\chi $ increases with increasing $T$ and finally
approaches zero under different electric field $Es$ and the magnetic field $
Bs$.

\section{Conclusions}

In summary, this paper reports on a theoretical study on the modulation of de Haas-van Alphen effect
in graphene by electric field. Three major findings emerge from the
study. First of all, we find that both magnetization and magnetic susceptibility are
modulated by the electric field. At the zero or finite temperature, both
magnetization and magnetic susceptibility are predicted to oscillate periodically as
a function of reciprocal field $1/B$. The $
dHvA$ oscillation period $\Delta (1/B)$ is derived analytically. It is also
discovered that as the parameter $\beta $ increases, the values of
magnetization and magnetic susceptibility finally increase to positive infinity or decrease to
negative infinity at the exotic point $\beta _{c}=1$. Besides, the analytical results
indicate that the $dHvA$ oscillation amplitude increases abruptly to infinity for zero
temperature at $\beta _{c}=1$, but eventually collapses at a finite temperature directly
leading to the vanishing of the de Haas-van Alphen effect. The "vanishing" is accounted for the anomalous effect of
the electric field on Landau level, which arises from the
instability of a relativistic quantum field vacuum. In addition, it is
established that the magnetic susceptibility depends on the electric and
magnetic fields, which suggests the graphene should be a non-linear magnetic
medium. These phenomena, not available in the standard 2D electron gas, are
deemed as the consequence of the relativistic type spectrum of low energy
electrons and holes in graphene.

\section{Acknowledgments}

We thank Dr D. Q. Liu, Q. Wang, X. H. Zhang, B. H. Gong, L. Xu, and H.
W. Chen for helpful discussions. This work is supported by the Cultivation 
Fund of the Key Scientific and Technical Innovation Project£¬Ministry of 
Education of China (NO 708082).

\bigskip

\end{document}